\documentclass[reprint,aps]{revtex4-2}
\usepackage{times}
\usepackage{graphicx}
\usepackage{amsmath}
\usepackage{amssymb}
\usepackage{graphicx}
\usepackage{lmodern}
\usepackage{wrapfig}
\usepackage{xcolor}
\usepackage{hyperref}
\hypersetup{
    colorlinks = true,
    linkcolor = blue,
    anchorcolor = blue,
    citecolor = magenta,
    filecolor = blue,
    urlcolor = blue
    }
\usepackage{mathrsfs}
\usepackage{color}

\begin{document}
\title{Enhanced Phonon Peak in Four-point Dynamic Susceptibility in the Supercooled Active Glass-forming Liquids}
\author{Subhodeep Dey}
\thanks{These authors contributed equally}
\author{Anoop Mutneja}
\thanks{These authors contributed equally}
\author{Smarajit Karmakar}
\email{smarajit@tifrh.res.in}
\affiliation{
Tata Institute of Fundamental Research, 
36/P, Gopanpally Village, Serilingampally Mandal,Ranga Reddy District, 
Hyderabad, 500046, Telangana, India }

\begin{abstract}
Active glassy systems can be thought of as simple model systems that imitate complex biological 
systems. Sometimes, it becomes crucial to estimate the amount of the activity present in such 
biological systems, such as predicting the progression rate of the cancer cells or the healing time of 
the wound. In this work, we study a model active glassy system to understand a possible 
quantification of the degree of activity from the collective, long-range phonon response in the 
system. We find that the four-point dynamic susceptibility, $\chi_4(t)$ at the phonon timescale, 
grows with increased activity. We then show how one can estimate the degree of the activity at such 
a small timescale by measuring the growth of $\chi_4(t)$ with changing activity. A detailed  
finite size analysis of this measurement, shows that the peak height of $\chi_4(t)$ at this phonon 
timescale increases strongly with increasing system size suggesting a possible existence of an 
intrinsic dynamic length scale that grows with increasing activity. Finally,  we show that this peak 
height is a unique function of effective activity across all system sizes,  serving as a possible 
parameter for characterizing the degree of activity in a system.
\end{abstract}

\maketitle 
\section{Introduction}
Active matter refers to a class of systems in which the constituent elements or particles 
consume internal energy to get propelled apart from their usual motion due to thermal 
fluctuations \cite{ActReviewMadan,Ramaswamy2010}. On the other hand, 
glassy liquids or supercooled liquids are the systems whose particles start to move 
collectively with decreasing temperature (increasing density) until they get kinetically 
trapped near their putative glass transition temperature (density) 
\cite{KDSAnnualReview,RFOT2,Book1,KDSROPP,Berthier2011,MCTRevDas}. 
The former is a non-equilibrium system that shows spectacular dynamical properties 
like large-scale ordering, flocks, swarms, etc., and is one of the current hot topics of
 research \cite{vicsek1995,activemct,berthier2013,saroj2018,szamel2016,activerfot,
 chaki2020,merrigan2020,bi2016,caprini2020,vijay2007,janssen2019}. Many biological 
 systems are shown to have dynamical properties similar to 
 the glass-forming liquids in the presence of active driving.  Thus studying of the physics 
of glasses under activity can shed important information regarding the 
dynamical properties of biologically relevant processes in nature.

Supercooled liquids are disordered systems that have been looked upon for a long 
time but remain one of the major unsolved problems in condensed matter physics. 
Some of the successful theories of glass transition include the mode-coupling theory 
(MCT) \cite{MCTRevDas}, random first order transition (RFOT) theory \cite{RFOT1,RFOT2}, 
etc. Viscosity or relaxation time of the liquid increases very rapidly with increasing
supercooling and in typical experiments, one defines the calorimetric glass transition 
temperature, $T_g$, as the temperature at which the relaxation time of the system 
becomes too large ($\sim 100s$). One of the other hallmarks of supercooled liquids 
is the existence of dynamic heterogeneity (DH), which 
refers to the presence of regions with a significant variation in their dynamical properties and a 
growing dynamical correlation length in the system while remaining structurally similar to normal 
liquids \cite{Book1,flenner2010}. The growing DH with increasing supercooling can be quantified using multipoint 
correlators like $\chi_4$, $g_{uu}$, etc., which are defined later.

The field of active glasses lies at conjunction of two fields - the fields of active matter 
and the glass transition. In recent years, the field of active glasses become very fascinating 
and important to study because of its ubiquitous presence in biological processes \cite{Poujade2007,CochetEscartin2014,Angelini2011,Park2015,Malinverno2017,
Garcia2015,Parry2014, Nishizawa2017,Zhou2009,kakkada2018,takatori2020motilityinduced}. 
The active systems including cellular monolayer \cite{Angelini2011,Garcia2015,Cerbino2021,jacques2015,vishwakarma2020}, 
bacterial colonies \cite{takatori2020motilityinduced}, model experimental models 
\cite{PhysRevLett.107.108303,Ni2013,PhysRevLett.112.220602,deseigne2010,klongvessa2019} also show collective 
dynamical properties in which the particles in the medium are dynamically correlated 
up to a correlation length scale, termed as dynamic heterogeneity length scale, $\xi_d$.
The simulations are also able to predict most of these observations \cite{flenner2016,mandal2020,berthier2013,PhysRevLett.112.220602}. 
It is indeed interesting to study these systems as they offer a plethora of new phenomena 
that are not present in the equilibrium systems without active forcing. Recently, it has been 
shown that active glasses are inherently different from their equilibrium counterpart in 
their dynamical response. In particular, these systems show strong growth of DH with  
increasing activity which can not be understood by an effective temperature like 
equilibrium theory \cite{Kallol}, while there are other suggestions \cite{Ashwin,cugliandolo2019}.

\begin{figure}[!thpb]
\includegraphics[width=0.48\textwidth]{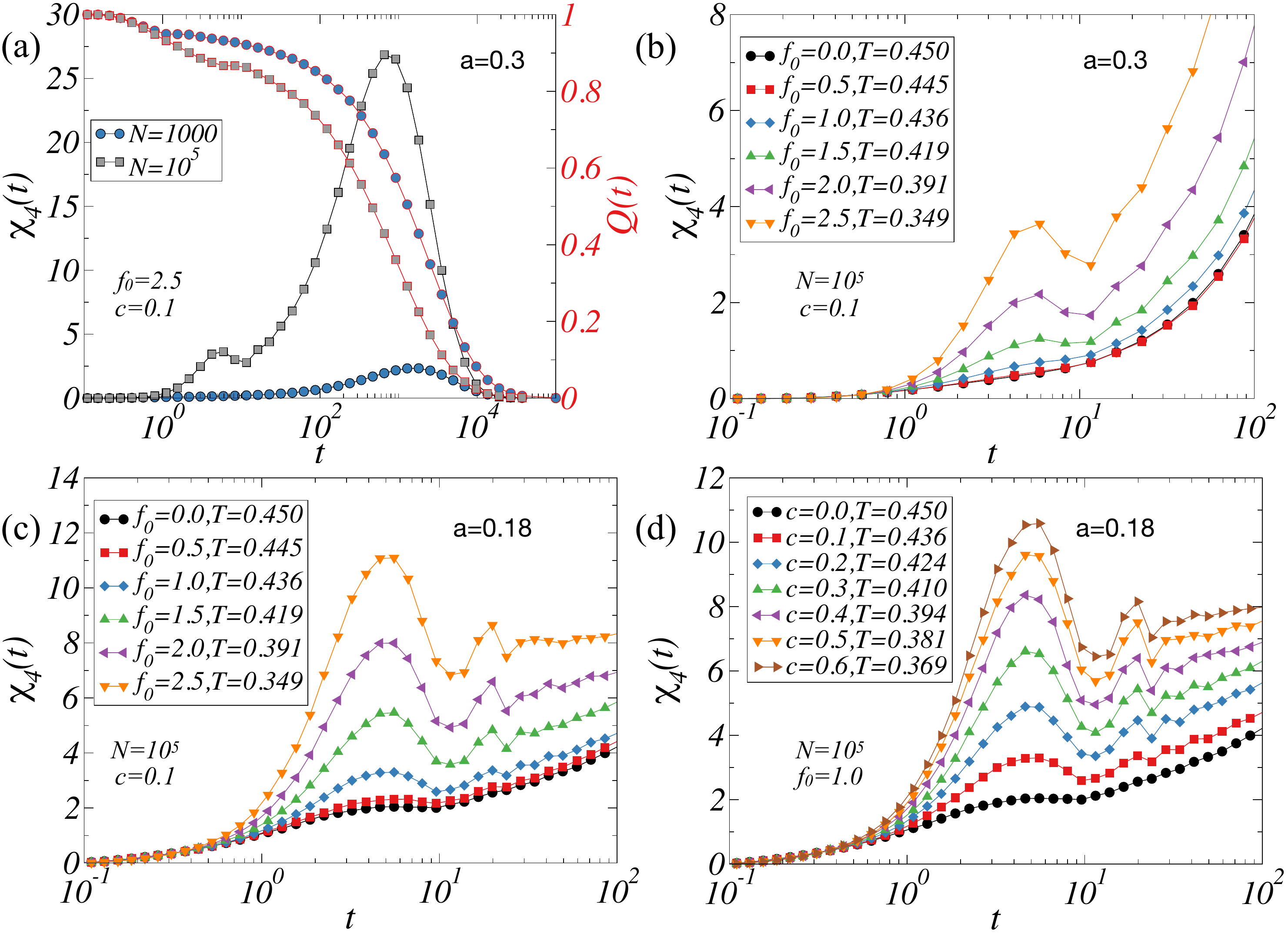}
\caption{(a). The overlap function $Q(t)$ (black), and its fluctuations $\chi_4(t)$ (red) is plotted against 
time in log scale for two different system sizes. It can be seen that the short 
time peak in $\chi_4(t)$ is more evident for larger system size.
(b). This panel highlights the systematic increase in the first peak height, $\chi_4^{P1}$, with 
increasing activity force, $f_0$. In panel (a) \& (b) analysis is done using the caging parameter $a=0.3$ in the Heaviside function.  
(c). The panel shows the same plots as (b) but with $a=0.18$, chosen to increase the peak signal even 
for smaller system sizes and smaller activities.  
(d). Similar analysis is done with changing $c$, while keeping active force $f_0=1.0$, constant.
Note that, the temperature is chosen to keep the $\tau_\alpha$ same for different activities. Also, this analysis is 
done on the subsystem with the linear length of $L/3$, where $L$ is the system length, to include the 
missing fluctuations and better averaging.
}
\label{Fig1}
\end{figure}

In this study, we have looked at the effects of active forcing on the dynamics of a model supercooled 
liquid at a timescale close to their vibrational timescale via extensive large-scale computer simulations. 
The phonons in a crystal are well defined because of the ordered structure, while in glasses, they are 
not because of the underlying disorder. Nonetheless,  long-wavelength dynamical correlated motion is 
also present in the deeply supercooled regime suggesting the vibrational motion of the system in deep 
potential energy minima. This vibrational motion of the system in the supercooled regime leads to the 
growth of a small peak in the four-point correlation function (Fig.\ref{Fig1}) ($\chi_4^{P1}=
\chi_4(t=t^*)$) at the short time $\beta$-relaxation regime \cite{SKPRL2016,Cohen2012}. The 
time at which the peak appears ($t^*$) is of the same order of magnitude as that of 
$\beta$-relaxation time, $\tau_\beta$. It was also highlighted in \cite{SKPRL2016} that the 
short time peak in $\chi_4^{P1}$ disappears if one does Monte Carlo Simulation or 
over-damped Brownian Dynamics simulation of the same model where the vibrational 
motion of the system will be entirely missing or suppressed. The phononic nature of 
$\chi_4^{P1}$ and the fact that $t^*$ is not the same as $\tau_\beta$ will be elucidated 
later in detail. We observed that $\chi_4^{P1}$ increases with increased 
activity in the system (Fig\ref{Fig1} (b-d)) even though the structural relaxation time 
is kept similar for various activities by choosing the temperature appropriately. These 
observations seem to suggest that the 
spatial extent of the collective modes probably extended further with increasing activity while the 
frequency of vibration remained the same. Thus, enhancement of the amplitude of phonon modes 
under active driving force can be a good quantifier for the degree of activity in the system. 
Thus this quantifier can be measured in experiments as measurement of the collective 
vibrational motions in various physical and biological systems would not be difficult because
the data acquisition will be of a shorter duration. 

On the other hand, the observed strong system size effect in $\chi_4^{P1}$ can be used 
to estimate an underlying 
intrinsic length scale of the system related to the activity. Tah et al. \cite{TahPRR} suggested the 
equivalence of the dynamic length scale at various time scales, including at $\tau_\beta$. Thus it is 
tempting to equate the length scale to that of the dynamic heterogeneity length scale, but one should 
be careful as the evolution of dynamic length scale, $\xi_d$ in active glassy systems may not be the 
same as the equilibrium behaviour. Further studies along that line are needed to draw a firm
conclusion. The rest of the paper is organized as follows. First, we will briefly 
discuss the model and methods, then show the phonon nature of the first peak along with systematic 
finite-size scaling and Block analysis (described later) to obtain the underlying length scale. 
This paper will end by 
discussing the role of effective activity parameters and their possible importance to future 
researches on active glasses.



\section{Models and methods}
In this work, in order to perform the finite-size scaling (FSS), we carry out molecular dynamics 
simulations over a range of system sizes ($N=400 - 100000$) of a binary glass-forming liquid in three 
dimensions (3d), well known in the literature by the name of Kob-Anderson model \cite{KA}. In the rest 
of the paper, it will be referred to as the 3dKA model. The model consists of larger (A-type) and 
smaller (B-type) particles in the ratio of $80:20$, and we randomly choose the `$c$' fraction of 
particles as active particles. We varied $c$ in the range $[ 0 - 0.6 ]$. To introduce the activity in the 
system, we choose the active particles to follow the run and tumble particle (RTP) model with variable 
active forces $f_0$ and fixed persistent time, $\tau_p=1.0$ in Lennard Jones units. The concentration 
of the active particles ($c$) is varied for a fixed $f_0$ in order to study the generic nature of the 
obtained results. It is important to note that the RTP model is crucial for this study. For example, one 
can use the active Brownian particle (ABP) model, but the most crucial drawback of the ABP model is 
that it does not include the effect of the inertial term in the equation of motion of the active particles 
by construction. Our results highlight that the inertial term carries crucial and significant information 
about the passive as well as the active system. This inertial term is responsible for the collective motion 
of the particles throughout the system, which is responsible for generating the system's intrinsic 
characteristic in terms of long-wavelength phonon-mode.

\section{Results}
  \begin{figure*}[!htpb]
\includegraphics[width=.99\textwidth]{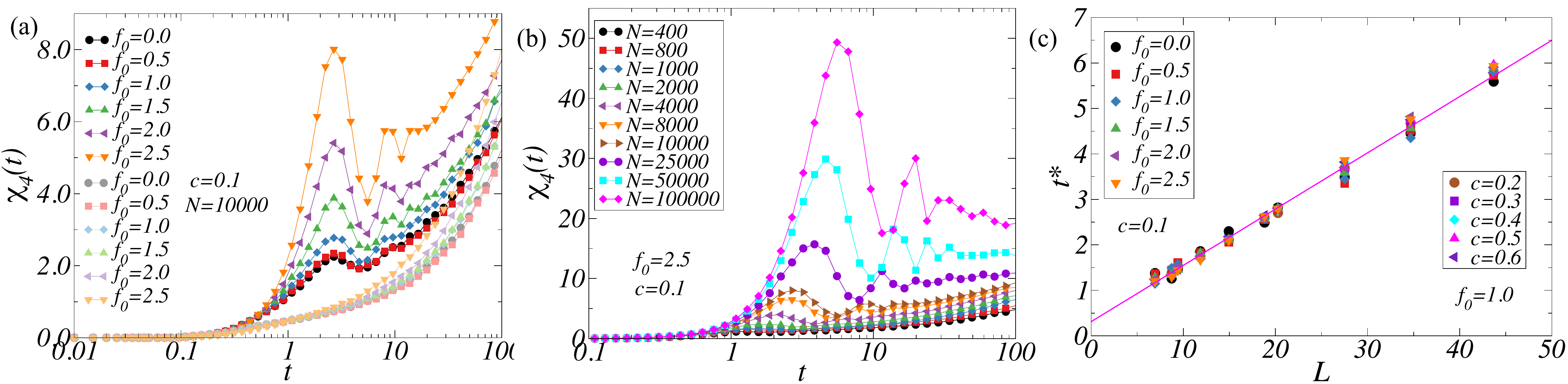}
\caption{(a)Small time behaviour of $\chi_4(t)$ is shown for active-3dKA system of $N=10000$ 
particles with increasing active force. The light colors represents the cage-relative $\chi_4(t)$, while 
the solid colors are the usual $\chi_4(t)$. The enhancement of the first peak with increasing activity  
can be clearly seen, while its collective nature is confirmed by the absence of the same in the cage 
relative quantity.
(b) System size dependence of the $\chi_4(t)$ is shown for the system with active forces $f_0=2.5$. 
The increase in peak height signifies the growth of underlying length scale.
(c) The peak position ($t^*$) as a function of the system size ($L$) is plotted. The variation turn out to be 
linear ($t^*\sim L$), implying again the phonon nature of the fluctuation.}
\label{Fig2}
\end{figure*}
The relaxation process in supercooled liquids is hallmarked with a small-time plateau in the overlap 
correlation function $Q(t)$ (see SI for definition), signifying the caging regime. $Q(t)$ indicates 
the fraction of particles still within their caging distance from the initial time. The caging 
distance can be parametrized by `$a$'. In this study, we choose the temperatures for different 
values of $f_0$ and $c$ such that the structural relaxation time $\tau_\alpha$ remains the 
same as shown in SI. $\tau_\alpha$ is defined as $Q(t=\tau_\alpha) = 1/e$. Fig.\ref{Fig1}(a) 
contains the $Q(t)$ and $\chi_4(t)$
plots for active systems with different system sizes.  This analysis of $Q(t)$ and $\chi_4(t)$ is done 
using the Heaviside function with parameter $a = 0.3$, the usual value used for the 3dKA model. On 
increasing the system size, one observes that the peak at early $\beta$-regime emerges and has a 
systematic trend with activity (see Fig.\ref{Fig1}(b) for $N = 10^5$). Note that the time at which the 
peak appears is linked to the long-range vibrational motion and hence depends on system size in a 
well-defined manner, as discussed in the subsequent paragraph. To further enhance the signal, we 
choose the parameter $a = 0.18$ in the rest of our analysis (see SI for details). 
Fig.\ref{Fig1}(c) shows the monotonic increase in the first peak height of $\chi_4^{P1}(t^*)$ 
for $N=10^5$. Also, the systematic increase in the peak height with increasing the concentration 
of the active particles (Fig.\ref{Fig1}(d)) again suggests the peak height maybe directly related 
to the degree of activity in the system and probably not related to the microscopic origin of 
the activity. Thus there exists a one-to-one relationship between the amount of activity in the 
system and the first peak height. This is indeed nice because $\chi_4^{P1}$ can be a direct
measure of the amount of activity in the system. Next, we discuss the phononic nature of the 
first peak of $\chi_4(t)$ and the characteristic timescale, $t^*$.

The confirmation of the phononic nature of the fluctuations comes from two facts. First, suppose this signal is powered by the collective motion of particles. In that case, one can suppress the signal by calculating the same quantities relative to the nearest neighbour cage \cite{Vivek1850,Illing1856,Mazoyer2009} (cage relative quantities) (see SI for definitions). In Fig.\ref{Fig2} (a), we highlight the small-time peak of $\chi_4(t)$ with $a = 0.18$ in the bolder colours, while the lighter colours are the cage relative $\chi_4(t)$ plots. One can see the absence of the same peak in the cage relative $\chi_4(t)$, confirming its origin from the collective motions. Note that subsequent peaks can be explained as coming due to the propagation of the same mode in the system. Now, one may argue that all collective motions need not be phononic in nature, so although this confirms the collective nature of the motion, it does not rule out other possibilities. Thus, we look at the system size dependence of the time when the peak appears ($t^*$), as shown in Fig.\ref{Fig2}(b). One can clearly see that the characteristic time increases with increasing system size. If it is due to a phonon, then one expects the characteristic time to scale linearly with system size according to the dispersion relation of phonon, $\omega  = C k$, where $\omega$ is the phonon frequency, $C$ is the sound speed, and $k$ is the wave vector. Fig.\ref{Fig2} (c) shows that indeed $t^*\sim L$, where $L$ is the linear dimension of the system. It is interesting to note that although $\chi_4^{P1}$ increases with increasing activity, the characteristic time scale does not seem to be dependent on the activity. This suggests that phonon mode's amplitude gets enhanced with increasing activity without much change in the phonon frequency. This fact surely warrants further investigations.

\begin{figure*}[!htpb]
\includegraphics[width=.98\textwidth]{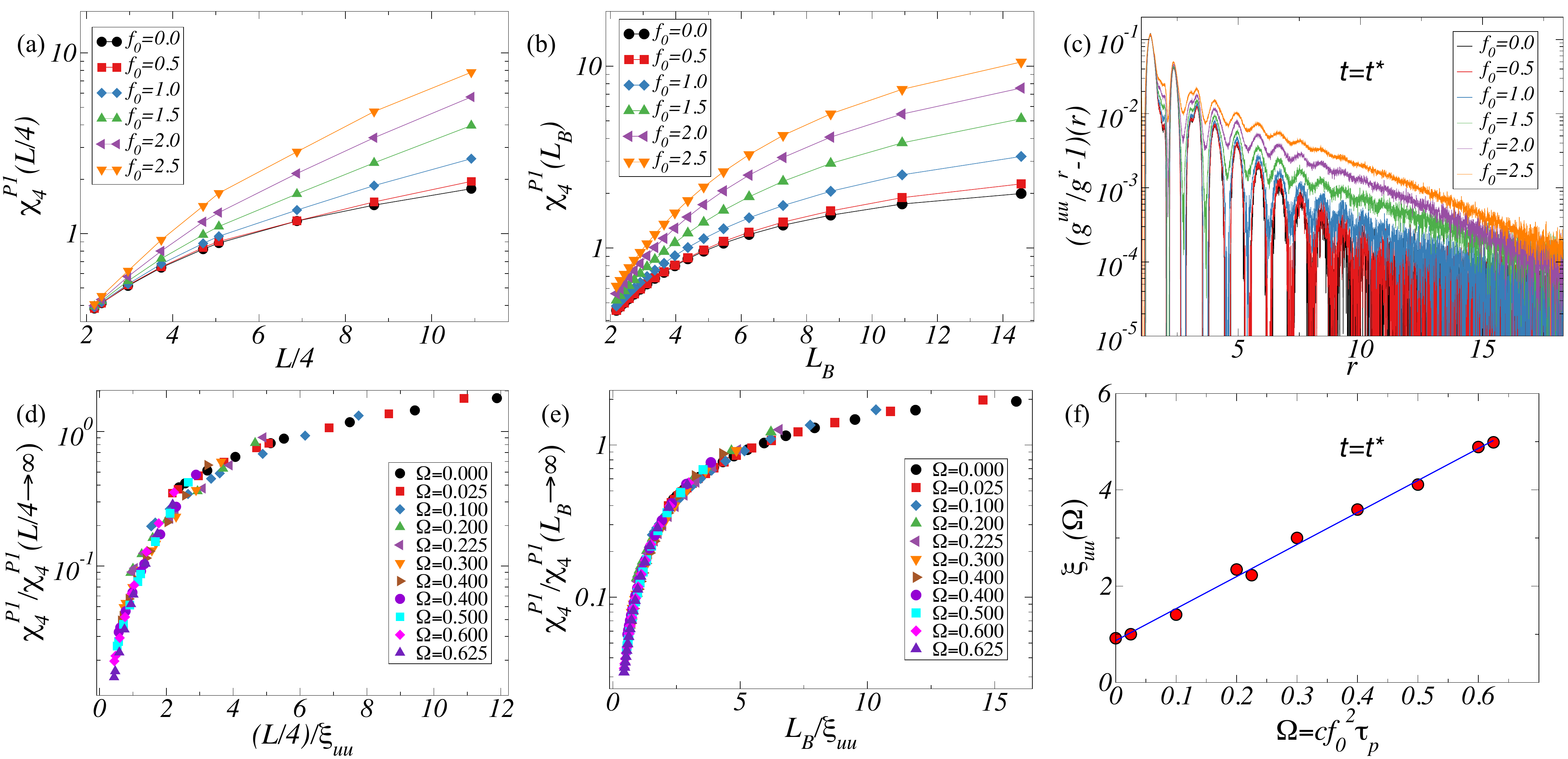}
\caption{(a). $\chi_4^{P1}$ is calculated and plotted against different system sizes, for various active forces to perform the finite size scaling. The calculation is done on subsystem with linear length of $L/4$ to include the missing fluctuations and to obtain better better averaging.  
(b). The peak height $\chi_4^{P1}$, is calculated for and plotted against the block lengths for different activities. The block analysis is performed for $N=10^5$ system. 
(c). The quantity $(g^{uu}-g^r)/g^r(r,\Delta t)$ is plotted as a function of spatial distance for $\Delta t=t^*$, here $g^{uu}(r,\Delta t)$ is the displacement-displacement correlation function (Eq.\ref{guu}), and $g^r$ is the pair correlation function. The plotted quantity would exponentially decays to zero providing us the correlation length $\xi_{uu}$, which can be clearly seen increasing with increasing activity. The length scale variation with increasing activity can be obtained as a integrated area of the plots and is shown in panel (f). The obtained values are rescaled according to exponential fitted value for $f_0=0.0$ to obtain $\xi_{uu}$ (see SI for details).
(d). \& (e). These panels show the scaling collapse done using scaling ansatz Eq.\ref{scalingAnsatz1} by using the length scales obtained from panel (f).}
\label{LengthScale}
\end{figure*}	
The next question that comes naturally is the similarity of the signal with the two-dimensional 
systems. Because of the Mermin-Wagner theorem \cite{MW}, such fluctuations grow 
immensely with the increasing system size and diverge in the infinite limit \cite{Shiba2016}. To explore the same, 
we took up the finite size analysis (FSS) of the active system. Fig.\ref{LengthScale}(a) shows the system 
size dependence of $\chi_4^{P1}$ calculated and averaged over each quarter of the system length 
($L_4$). The quarter of the system length is used to increase the averaging and include many 
important missing  fluctuations like fluctuation in density, temperature, concentration of the active 
particles, etc. \cite{Block}. This method of finite size scaling analysis is known as `block analysis'
and henceforth we will use the same name in the rest of the article. The plot shows that the peak height increases rapidly with increased system 
size and would saturate for a large enough system size in contrast to 2D systems. Also, the increase in the peak value 
becomes more and more drastic with increasing activity. In Ref.\cite{Kallol} it has been shown 
that the dynamic correlation grows spatially with increasing activity. Our observation of finite 
size effects on the phonon peak also suggests the growth of some inherent dynamic length 
scale with increasing activity. Note that in Ref.\cite{TahPRR} it has been shown 
that the dynamic correlation length remains the same at various time scales, including the 
early $\beta$-regime. Thus it will be important to compute the dynamic length scale at the time scale 
of phonon peak in an independent manner to see whether the observed finite-size effects
in $\chi_4^{P1}$ can be rationalized using that length scale. To do so, we turn to compute  
the displacement-displacement correlation function at $t^*$ {\it i.e.} $g^{uu}(r,t^*)$ \cite{TahPRR,Poole1998}. It is 
defined as,
\begin{equation}
g^{uu}(r,\Delta t)=\frac{\left\langle  \sum\limits_{i,j=1,j\neq i}^Nu_i(0,\Delta t)u_j(0,\Delta t)\delta(r-|\textbf{r}_{ij}(0)|)\right\rangle}{4\pi r^2\Delta rN\rho \langle u(\Delta t)\rangle^2}
\label{guu}
\end{equation}  
 where, $u_i(t,\Delta t)=|\textbf{r}_i(t+\Delta t)-\textbf{r}_i(t)|$, and $\langle u^2(\Delta t)\rangle=\langle\frac{1}{N}\sum_{i=1}^N\textbf{u}_i(t,\Delta t).\textbf{u}_i(t,\Delta t)\rangle$. $g^{uu}(r,\Delta t)$ is calculated at time $\Delta t=t^*$, along with the usual pair correlation function $g^r$.
The quantity $g^{uu}(r,\Delta t)/g^r-1.0$ would decay to zero as a function of $r$, providing 
the decorrelation length of particle's displacement over the time interval $\Delta t$. If one assumes 
the decay to be exponential, then the area under the curve would provide us with the correlation 
length without involving any fitting procedure.  Fig.\ref{LengthScale}(e) contains the semi-log plot of $g^{uu}(r,\Delta t)/g^r-1.0$ calculated 
for system size of $N=10^5$. The observed near linear behaviour confirms the exponential decay. 
The obtained length scale ($\xi_{uu}$) with increasing activity in the system is plotted in 
Fig.\ref{LengthScale}(f). The values in the plot are the integrated areas scaled with the 
absolute value of the length scale obtained by fitting an exponential function to the data 
for the passive case, $f_0=0.0$. Note that the length scale grows by a factor of $\sim 5.5$, 
while the structural relaxation times  remain similar. We then performed the FSS of the data 
presented in Fig.\ref{LengthScale}(a) using
the same  length scale using the following scaling ansatz 
\begin{equation}
\chi_4^{P1}(L,\Omega) = \chi_4^{P1}(L\to\infty,\Omega) \mathcal{G}\left[\frac{L}{\xi_{uu}(\Omega)}\right],
\label{scalingAnsatz1}
\end{equation}
where, $\chi_4^{P1}(L\to\infty,\Omega)$ is the large system size asymptotic value of the 
susceptibility peak and $\Omega = cf_0^2\tau_p$ is the effective activity in the system as 
suggested in Ref.\cite{Mandal2016}.
\begin{figure*}[!htpb]
\includegraphics[width=.93\textwidth]{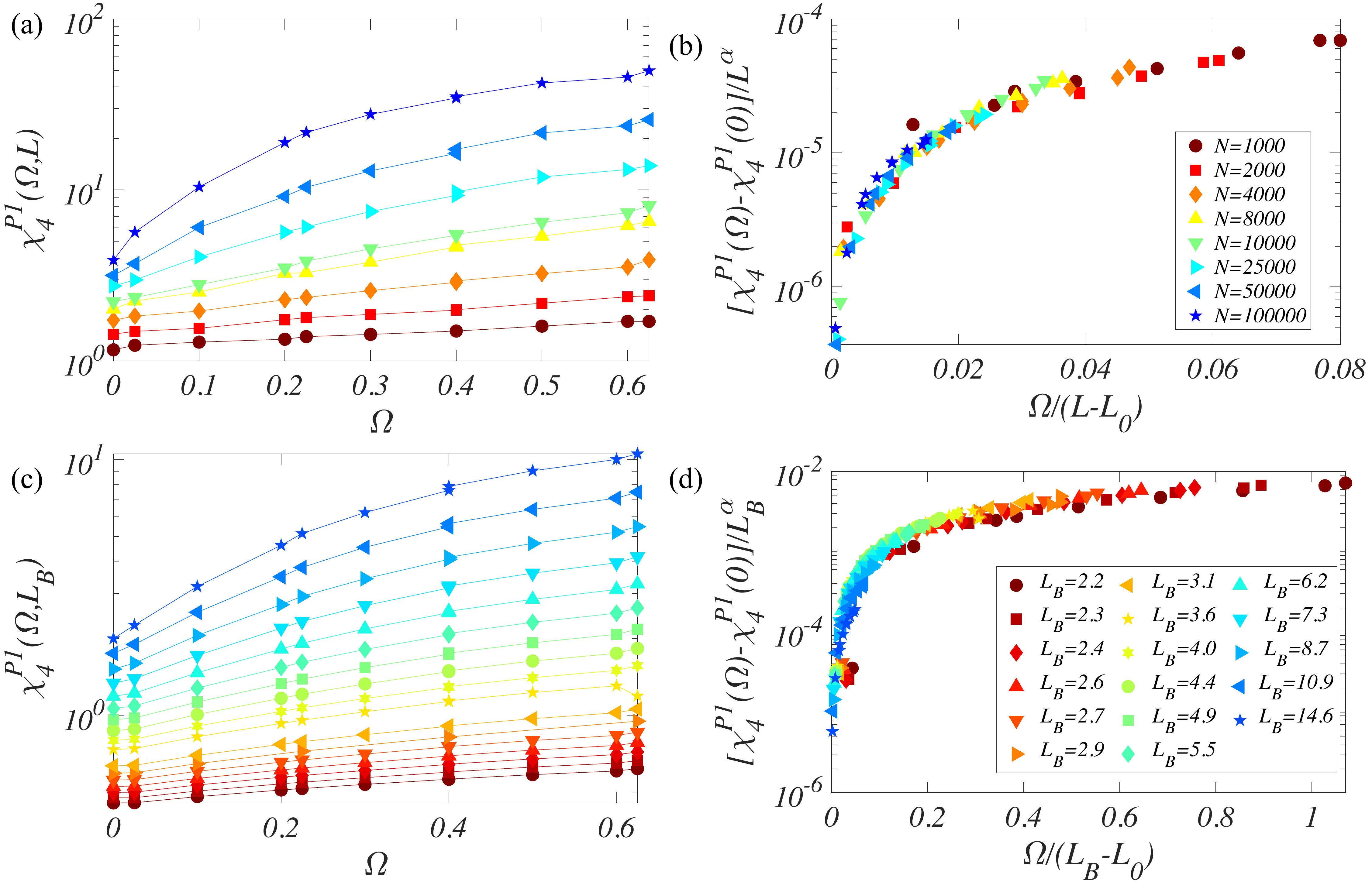}
\caption{(a) The peak height, $\chi_4^{P1}$ is plotted against the effective activity parameter $\Omega=cf_0^2\tau_p$ for different system sizes and different block lengths (c). The singular function while changing the $c$ and $f_0$ in $\Omega$ for a particular system size or block length suggests the one-to-one correspondence of $\chi_4(PI)$ and net activity in the system. The scaling anstaz Eq.\ref{scalingAnsatz} is used with the parameters, $L_0=1.6$, and $\alpha=4.0$ to collapse both the finite size data and block analysis data in panels (b) and (d) respectively.}
\label{ActParam}
\end{figure*}
The data collapse using Eq.\ref{scalingAnsatz1}  observed in Fig.\ref{LengthScale}(b) is indeed very good, suggesting that the length scale can explain the observed finite size effect elegantly. Next, we performed block analysis of the peak height $\chi_4(L_B, f0, t = t^*)$ computed for varying block size, $L_B$ as shown in Fig.\ref{LengthScale}(c), to reconfirm the connection of the peak and underlying dynamic correlation. To our expectation, the block length ($L_B$) variation of $\chi_4^{P1}$ can be collapsed by assuming the same length scale $\xi_{uu}(\Omega)$ as shown in Fig.\ref{LengthScale}(d). The details of the block analysis are given in the SI. Note that in the top panel of Fig.\ref{LengthScale}, we have shown data with different $f_0$ and not included the data with different $c$ for better clarity. While in the bottom panel, all of the variations in terms of effective activity $\Omega$ (also discussed later in the text) are included. Note that $\chi_4^{P1}$ depends on the caging parameter $a$, which we have chosen to be $a = 0.18$, while the scale obtained from the displacement-displacement correlation function does not depend on such parameters. So one can conclude that the correlated dynamics extend further into the space with increasing activity even at the vibrational time scales, and the increasing peak height signifies the enhancement of phonon amplitude. Thus, $\chi_4^{P1}$ indeed seems to be a good and direct measure of both the activity in the system and the correlated dynamic length scale.

Activity in the biological or model systems can be present in various forms. We also tried to reciprocate 
the same in our system in the following three ways: by changing the concentration of active particles, 
$c$, or by changing the magnitude of the force on those particles, $f_0$, or by increasing the 
persistence time of active particles, $\tau_p$. In Ref.\cite{Mandal2016}, the variable $
\Omega=cf_0^2\tau_p$ was used to quantify net activity in the system, and it was shown that within a 
range of parameter values, this parameter uniquely defines the degree of activity in the system. This 
means that if one changes $f_0$, $c$, and $\tau_p$, keeping $\Omega$ the same, one would expect 
the system's dynamical behaviour to be the same. It would be interesting to check if $\chi_4^{P1}$ also 
follows this behaviour with changes in the parameters across system sizes. 
Fig. \ref{ActParam}(a) 
shows the plot of $\chi_4^{P1}$ with respect to $\Omega$, and one can see that for both the change 
in concentration, $c$, and the change in the active force, $f_0$, the $\chi_4^{P1}$ follows a universal 
function for given system size. The dependence of $\chi_4^{P1}$ on $\Omega$ has strong system size 
effects. To understand the same in a unified manner, we develop a scaling theory as follows: For a 
given system size with linear dimension, $L$, the $\chi_4^{P1}$ seems to show a saturation tendency 
above a certain activity, $\Omega = \Omega^*(L)$. It is also evident from the data that $\Omega^*(L)$ 
seems to increase with increasing systems size. We proposed the following scaling function to 
rationalize the observation,
		
\begin{equation}
\chi_4^{P1}(\Omega, L) = L^{\alpha}\mathcal{F}\left( \frac{\Omega}{\Omega^*}\right) + \chi_4^{P1}(0,L).
\label{scalingAnsatz}
\end{equation}
Note that at $\Omega = 0$, $\chi_4^{P1}$ has a finite value that depends on the system size, so we 
subtract out the part of the contribution in $\chi_4^{P1}$, which does not come due to activity. Now, if 
one takes into account that there is a growing correlation in the system due to activity, then at certain 
activity $\Omega  = \Omega^*$ , the correlation length $\xi_{uu}(\Omega^*)$ will be similar to the 
length of the simulation box, $L$. If one then increases $\Omega$ further, the correlation length will then 
be bounded by the finite size of the simulation box. The corresponding susceptibility will also saturate 
to a value solely controlled by the system size. If one now assumes a dynamical scaling behaviour 
similar to critical phenomena, then one can expect $\chi_4^{P1}(\Omega \to \infty,L) \sim L^{\alpha}$, 
$\alpha$ being one of the scaling exponents. Now, on the other hand, $\Omega^*$ can be obtained by 
demanding $\xi_{uu}(\Omega^*) \sim \zeta L$, where $\zeta$ is a scale factor of order unity. Note 
that $\xi_{uu}(\Omega)$ shows a linear dependence with $\Omega$, as shown in Fig.\ref{LengthScale}
(f). If we assume $\xi_{uu}(\Omega) = \xi_{uu}(0) + A\Omega$, then $\Omega^* \sim L - L_0$ up to 
an overall scale factor, where $L_0$ is a scaling parameter that depends on the correlation length at $
\Omega = 0$. Suppose these scaling arguments indeed capture the underlying physics. In that case, 
one expects that all data shown in Fig.\ref{ActParam}(a) will fall on a master curve if  $[\chi_4^{P1}
(\Omega, L) - \chi_4^{P1}(0, L)]/L^{\alpha}$ is plotted as a function of scaled frequency $\Omega/(L - 
L_0)$ with an appropriate choice of the parameter $\alpha$ and $L_0$. The validity of this assumption 
is shown in Fig.\ref{ActParam}(b). The data collapse with $\alpha = 4$ and $L_0 = 1.6$ looks 
reasonable, suggesting that a scaling theory can describe both activity dependence and system size 
dependence of $\chi_4^{P1}$ in a unified manner. 

This also gives us the possibility to quantify the 
changes in the degree of activity in the system compared to its zero activity value by computing the 
first peak in four-point susceptibility. It will surely have advantages in experiments in which often 
estimating the degree of activity is not easy because activity often arises from the internal activity of 
the constituent particles that cannot be directly controlled in a precise manner by external means. 
In particular, in 
experiments involving imaging techniques, often a small part of the whole system is looked at. In that 
context, it will be essential to check the validity of the same scaling theory if one studies the variation 
of $\chi_4^{P1}$ using the block analysis method. In Fig.\ref{ActParam}(c) \& (d), we did the same 
analysis for $\chi_4^{P1}$ computed for block sizes ($L_B$) at various activities ($\Omega$), and the 
scaling collapse is obtained using the same parameters as used before. The data collapse was again 
observed to be good. This gives us the confidence that this method will be very useful in experiments.

\section{Conclusions}
To conclude, we have shown that with increased activity, the fluctuations in the relaxation process in 
the $\beta$-relaxation regime increase systematically, which was then shown to be linked with the 
cooperative motion of the particles. This leads to an important inference that one can obtain the amount 
of activity in the system by looking at its vibrational relaxation process and the associated four-point 
dynamic susceptibility, $\chi_4(t)$. This in the future might play an essential role in determining the 
degree of activity in experimental systems where the source of active driving can come from the 
internal processes of the constituent particles and direct control and estimation of the total activity in 
the system might not be immediately available. In particular, there are experimental studies that 
measured the $\chi_4(t)$ in systems like epithelial monolayers \cite{Malinverno2017}, cell 
assemblies \cite{Cerbino2021}. With the proposed method, one will be able to obtain valuable 
information about the net activity in the system and a growing dynamical correlation length
even by studying the short time dynamics which requires shorter data accusation. This we hope
will surely encourage many future experiments both in biological systems as well as synthetic
active matter system. Finally, we provided a scaling theory to understand the 
activity dependence as well as the system size dependence of four-point susceptibility in 
a unified manner. The results clearly show that the peak height of $\chi_4(t)$ at a short 
timescale is probably be a function of the effective activity parameter $\Omega = cf_0^2\tau_p$ 
at least within the studied system sizes and parameter ranges.
    
\begin{acknowledgments}
SK would like to acknowledge funding by intramural funds at TIFR Hyderabad 
from the Department of Atomic Energy (DAE) under Project Identification 
No. RTI 4007. Core Research Grant CRG/2019/005373 from Science and 
Engineering Research Board (SERB) as well as Swarna Jayanti Fellowship grants 
DST/SJF/PSA-01/2018-19 and SB/SFJ/2019-20/05 are acknowledged.
\end{acknowledgments}

\bibliographystyle{apsrev4-2}
\bibliography{ActPhonon} 
\end{document}


\title{Enhanced Phonon Peak in Four-point Dynamic Susceptibility in the Supercooled Active Glass-forming Liquids - Supplimentary Information}
\author{Subhodeep Dey}
\thanks{These authors contributed equally}
\author{Anoop Mutneja}
\thanks{These authors contributed equally}
\author{Smarajit Karmakar}
\email{smarajit@tifrh.res.in}
\affiliation{
Tata Institute of Fundamental Research, 
36/P, Gopanpally Village, Serilingampally Mandal,Ranga Reddy District, 
Hyderabad, 500046, Telangana, India }

\maketitle 

\section{Models and methods}
In this work, we have performed extensive molecular dynamics simulation of Binary mixture of 
Lennard-Jones (BMLJ) particles interacting via the following potential. 
The potential is smoothed such that  $2^{nd}$ derivative of the 
potential will be continuous at the cut off radius $r_c$,
\begin{equation}
 \phi(r)=\begin{cases}
 4\epsilon_{\alpha\beta} \left[\left(\frac{\sigma_{\alpha\beta}}{r}\right)^{12}-\left(\frac{\sigma_{\alpha\beta}}{r}\right)^{6}+c_0 + c_2 r^2\right]    &,r<r_c  \\
 0     &,r\geq r_c
 \end{cases}
\end{equation}
Here, $\alpha$ and $\beta$ refers to large ($A$-type) or small ($B$-type) particles 
respectively. The ratio of $A:B = 80:20$ is maintained. This model is well known in the 
literature as the Kob-Andersen model (3dKA) \cite{KA}. The interaction strengths and particle
diameters are $\epsilon_{AA}=1.0$, $\sigma_{AA}=1.0$, $\epsilon_{AB}=1.5$ , $\sigma_{AB}=0.8$, 
and $\epsilon_{BB}=0.5$, $\sigma_{BB}=0.88$, and $r_c=2.5\sigma_{AB}$. The number 
density ($\rho$) of the system is taken to be $1.2$ for all the simulations. The units of 
length, energy and time are given by $\sigma_{AA}$, $\epsilon_{AA}$ and 
$\sqrt{\frac{\sigma_{AA}^2}{\epsilon_{AA}}}$ respectively. The integration step size is 
chosen to be $\delta t = 0.005$ for all our simulations.

\subsection{Introducing Activity: Run and Tumble particle model (RTP)} 
Activity in this study is introduced in the system by randomly choosing the $c$ fraction of particles as active particles. The active particles get an extra active force $f_0$ along any random direction while keeping zero vector sum of total active forces. The direction of the active force changes after the persistent time $\tau_p$. The active force on $i^{th}$ particle reads as,

\begin{equation}
F^A_i=f_0 \left(k^i_x \hat{x} +k^i_y \hat{y} +k^i_z\hat{z} \right),
\end{equation}

where $k^i_x$, $k^i_y$, $k^i_z$ are randomly chosen from $\pm1$, after every persistent time interval. Thus an 
active particle can have one of the eight possible directions. Also, to maintain the momentum conservation of the 
system, there should be an even number of active particles in the system with total active force equated to zero, 
which mathematically implies, $\sum_{\alpha,i} k^i_\alpha=0$. Thus, the total activity in our system is defined by 
three parameters, the active force magnitude $f_0$, the concentration of active particles $c$, and the persistent 
time $\tau_p$. In this study, we have first varied $f_0$ in the range $f_0 \in [0.0 - 2.5]$ while keeping $c=0.1$ 
and $\tau_p=1.0$ constant. Then the concentration $c$ is varied in range $c \in [0.0 - 0.6]$ while keeping 
$f_0=1.0$ and $\tau_p=1.0$. Note that we haven't changed the persistent time in this study. As large persistent 
time leads to a complete dynamical behaviour of the system as reported in \cite{mandal2020}, we kept it small 
and fixed to study the effect of activity in the glassy regime only.
\section{Thermostat}
The thermostat is one of the main challenges in non-equilibrium simulations. In particular, it seems that various 
thermostats fail to maintain a constant temperature in the presence of active forces. Thus, we have used the 
three-chain Nos\'e-Hoover thermostat \cite{Allen} to get the desired temperature which is known to maintain true canonical 
ensemble fluctuations in equilibrium. The relaxation time of the thermostat is set to $10-20$ times the 
simulation time-step. We also checked another thermostat known as Gaussian thermostat \cite{zhang1997}, which is also 
found to be able to control the temperature well in the presence of activity. The results obtained using these two 
thermostats are quantitatively similar.
\section{Overlap correlation function, $Q(t)$}
To characterize the system's dynamical properties, we have computed the two-point density-density correlation 
function of the system. For simplicity, we have computed the overlap correlation function $Q(t)$, defined as 
\begin{equation}
 Q(t) = \frac{1}{N}\sum_{i=1}^N w(|\vec{r}_i(0)-\vec{r}_i(t)|),
 \end{equation}
where $w(x)$ is a window function, and it is one if $x < a$, where $a$ is a parameter that is chosen to remove 
the possible initial decorrelation that can happen due to the fast vibrational motion of the particles. $\vec{r}_i(t)$ 
is the position vector of particle $i$. The value of `a' is typically chosen from the plateau region in the system's 
`mean-square displacement (MSD)'. In the supercooled liquid regime, the MSD shows a plateau representing the 
cage exploration of the system during the transition of the particle dynamics from the ballistic to diffusion region. 
One often chooses this value to maximize the signal strength of the fluctuations of $Q(t)$, which is defined later 
as $\chi_4(t)$. We will discuss this in detail in the subsequent paragraph. Typically, the value of `a' is chosen to 
be $0.3$. This relaxation time, $\tau_{\alpha}$, is obtained
as $\left<Q(t=\tau_{\alpha})\right>=e^{-1}$ where $\langle \cdots \rangle$ refers to ensemble average. The 
system is equilibrated long enough (typically $\sim 50 \tau_\alpha$) so that the system's dynamic is ergodic in 
nature. We did further $100\tau_\alpha$ long runs for gathering
data. We averaged our data over $32$ statistically independent ensemble runs for all systems $N \le 10000$ and 
$10$ simulations runs for $N > 10000$ respectively.
      
\section{Four-point correlation function, $\chi_4(t)$}
Four-point correlation susceptibility, $\chi_4(t)$ is the measure of the fluctuation in two-point 
correlation function $Q(t)$. It is defined as 
\begin{equation}
 \chi_4(t) = N \left[\langle Q(t)^2\rangle - \langle Q(t)\rangle^2\right].
\end{equation}
We averaged $\chi_4(t)$ over $32$ ensembles for simulations with $N \le 10000$ particles
and $10$ ensembles for $N > 10000$. 
    
Note that $\chi_4(t)$ is one of the best ways to characterize the degree of heterogeneity in a system. This 
typically quantifies the sizes of different regions with fast and slow dynamics. The time at which $\chi_4(t)$ peaks 
is close to the relaxation time $\tau_{\alpha}$ that is $\chi_4(t=\tau_\alpha) \simeq \chi_4^p$. The increasing 
system size shows one more peak at shorter timescale around the $\beta$-relaxation regime. It is found that 
peak at short timescale can be enhanced by a suitable choice of the cut-off parameter `a'. For $a = 0.3$, most of 
the small-amplitude motion of the particle is masked, which is important to pick up the long-wavelength mode at 
low temperatures. To enhanced the peak height of  $\chi_4(t)$ at short timescale we have chosen $a = 0.18$. The 
peak at short timescale is defined as $\chi_4^{P1}$ and the time corresponding to the first peak (maxima) of $
\chi_4(t)$ as $t^*$.

\section{Cage-relative Displacement}
To separate out the collective behaviour that may arise from the vibrational dynamics, especially at a short 
timescale, we have computed the cage-relative (CR) displacement of the individual particles, which is defined as $
\vec{r}_{i,CR}(t)$.
\begin{align}
 \vec{r}_{i,CR}(t) = [\vec{r}_i(t) - (\vec{r}_{i,nn}(t) - \vec{r}_{i,nn}(0))]
\end{align}
where, $\vec{r}_{i,nn}(t)$ is  center of mass position of $N_{nn}$ nearest neighbours 
(nn) at time $t$ and it is defined as,
\begin{align}
 \vec{r}_{i,nn}(t) = \frac{1}{N_{nn}} \sum_{j=1}^{N_{nn}} [ \vec{r}_j(t) - \vec{r}_j(0) ].
\end{align}
Here we have used the cut-off value $r_c^{nn} = 1.3$ at the initial time to get the $N_{nn}$ number of nearest 
neighbours and then track those neighbour particle's motion with respect to that time origin. This modified cage-
relative displacement quantity has then been used to compute both cage-relative $Q(t)$ and $\chi_4(t)$.

\section{Block Analysis}
In this work, we have done extensive finite-size scaling analysis using the Block analysis method 
\cite{Block}. In this method, the whole system is divided into smaller subsystems, and then one studies all 
the above-mentioned correlation functions to incorporate some of the important fluctuations. For example, $
\chi_4(t)$ will have contributions coming from a number of particle fluctuations, density fluctuations, the 
concentration of particle species fluctuations, temperature fluctuations, etc. This is also one of the most natural 
ensembles, especially in experiments in which a subsystem is typically probed using various imaging methods. 
The two-point correlation for one subsystem can be redefined similarly as, 
\begin{align}
 Q(L_B,t)= \frac{1}{n_i}\sum_{j=1}^{n_i} [w(r_j(t)-r_j(0))],
\end{align}
where $L_B= (N/N_B)^{1/3}$ and, $N_B$ is the number of subsystems referred henceforth as blocks. $n_i$ is the 
number of particles in the block with level $i$. Now the average correlation of the function will be just
\begin{align}
\left<Q(L_B,t)\right> &=\frac{1}{N_B}  \sum _{i=1}^{N_B} Q(L_B,t).
\end{align}
Similarly, the four-point susceptibility for each block can be written as
\begin{align}
 \chi_4^{\prime}(L_B,t)&= [\left<Q(L_B,t)^2\right>-\left<Q(L_B,t)\right>^2].
\end{align}
So the averaged susceptibility will be given by
\begin{align}
\chi_4(L_B,t) &=\frac{N}{N_B} \sum _{i=1}^{N_B} \chi_4^{\prime}(L_B,t).
\end{align}
In this case $\left< \cdots \right>$ denotes averages over different grand canonical ensembles 
of size $L_B$.

\section{Choice of temperature for different activity}
\begin{figure*}[!htpb]
\begin{center}
\includegraphics[width=0.87\textwidth]{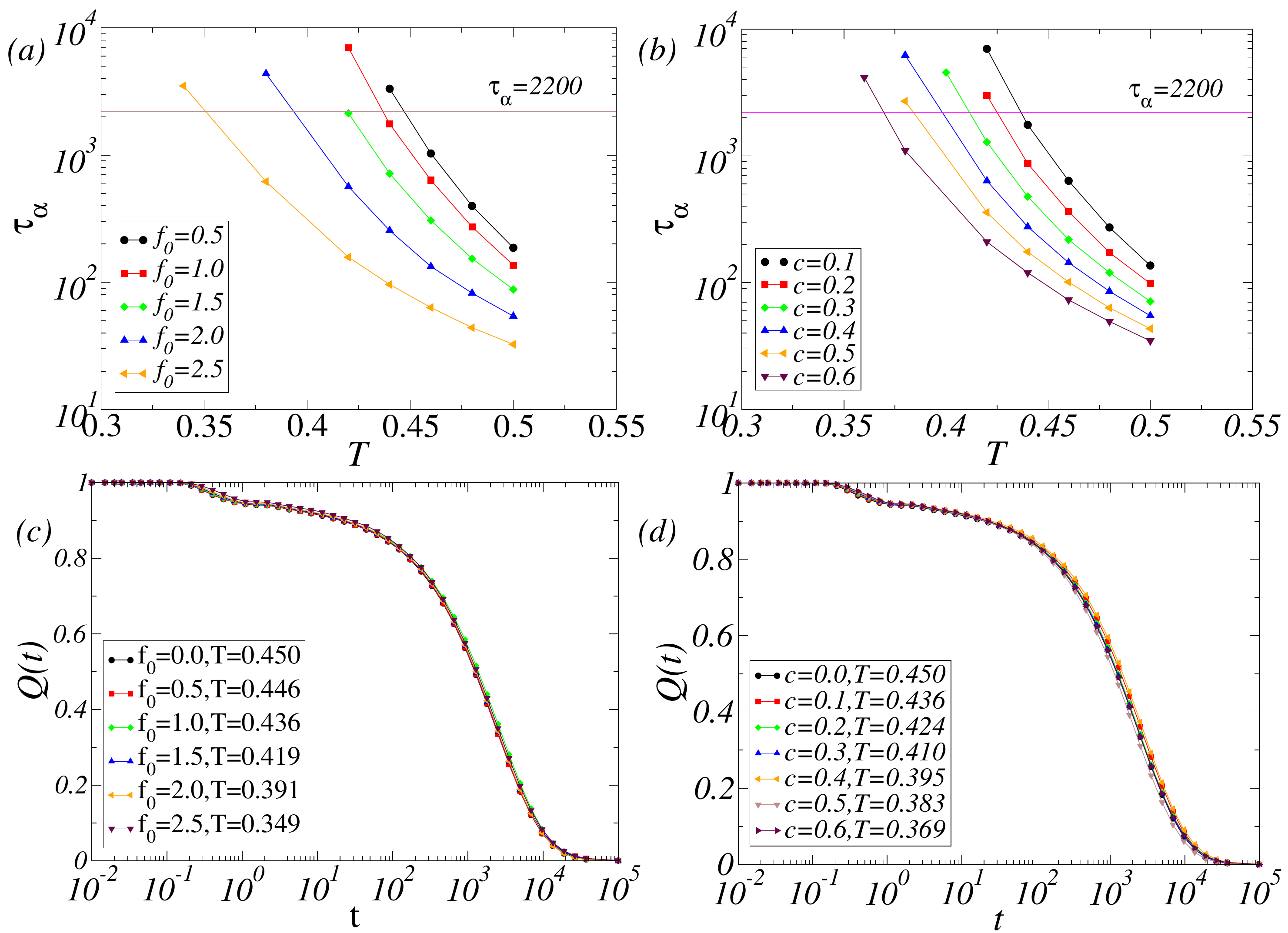}
\caption{(a). Relaxation time $\tau_{\alpha}$ vs temperature $T$ for different 
driving $f_0$ with fixed concentration of active particles $c = 0.1$ and persistent time 
$\tau_p=1.0$. Here the system size $N = 1000$. (b). Similar results for 
varying concentration at fixed value of $f_0=1.0$ and $\tau_p = 1.0$. 
(c). Two-point correlation function $Q(t)$ vs time $t$ . Here the temperature is 
chosen such that the relaxation time ($\tau_\alpha$) for different activity ($f_0$) is 
around $2200$. (d). Similar plots when we varied the concentration
of active particles $c$ keeping $f_0 = 1.0$ and $\tau_p$ constant. }
\label{tauAlphavsT}
\end{center}
\end{figure*}
To compare the effect of different activity we have fixed the relaxation time $\tau_{\alpha}$ for all the systems. 
For which we have looked at the temperature dependence of $\tau_{\alpha}$ for different value of activity 
parameter $\Omega = cf_0^2\tau_p$. Here $c$ is the concentration of the active particles, $f_0$ is the 
magnitude of the applied active force, and $\tau_p$ is the persistent time over which the directions of active 
forces change randomly. The change of $\tau_{\alpha}$ for different $T$ can be fitted well via VFT 
(Vogel-Fulcher-Tammann) fitting function (see top panels of Fig.\ref{tauAlphavsT})

\begin{align}
\tau_{\alpha}=\tau_0\exp[A/(T-T_0)].
\end{align}
By using the above fitting equation, we can find the temperature corresponding to a fixed relaxation time $
\tau_{\alpha}$ for a different activity. Here we have fixed the relaxation time $\tau_{\alpha}$ corresponding to $T 
= 0.45$ of a passive system, which is around $\tau_\alpha \sim 2200$ for $N = 1000$ particles. Firstly, the 
changes because of $f_0 = 0.0, 0.5, 1.0, 1.5, 2.0 ,2.5$ has been studied for $c=0.1$, and, $\tau_p=1.0$, as 
shown in top left panel of Fig.\ref{tauAlphavsT} and then the changes because of $c=0.0, 0.1, 0.2, 0.3, 0.4, 0.5, 
0.6$ are studied for fixed $f_0=1.0$, and, $\tau_p=1.0$ as shown in the top right panel of Fig.\ref{tauAlphavsT}. 
Subsequently, we choose the temperatures so that the relaxation time is the same across various activities for $N 
= 1000$. The corresponding correlation functions for these temperatures are shown in the bottom panels of 
Fig.\ref{tauAlphavsT}. One can see that the two-point correlation function falls on top of each other. The value of 
$a$ is $0.3$ in this case.

\section{First peak of dynamic susceptibility}
As discussed in the main article,  we set the value of $a = 0.18$ while studying 
$\chi_4(t)$ especially at short timescale. This allowed us to pick the first maximum of 
dynamic heterogeneity curve in the early-beta region very effectively. 
\begin{figure*}[!htpb]
\begin{center}
\includegraphics[width= 0.90\textwidth]{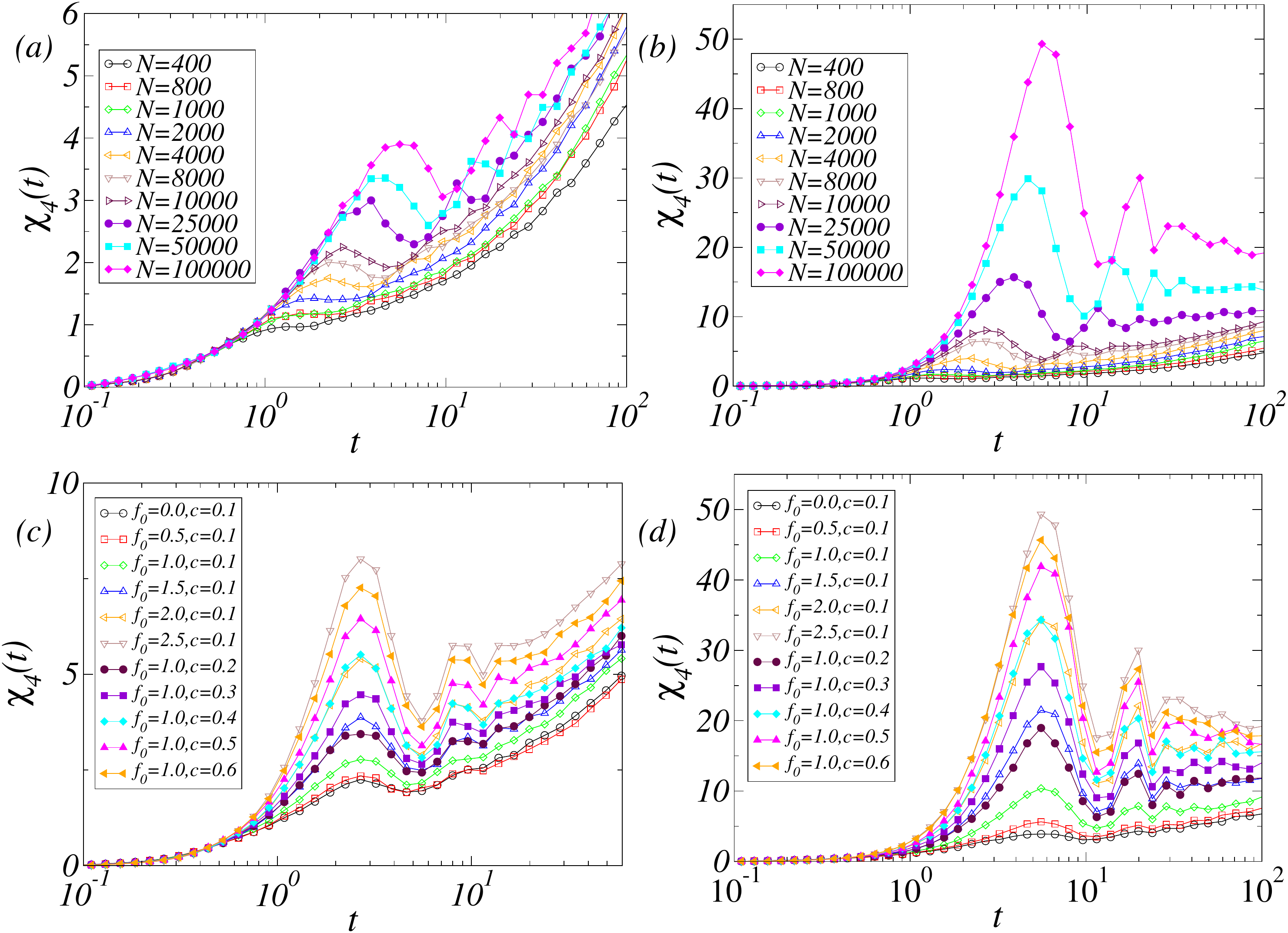}
\caption{(a). Dynamic susceptibility $\chi_4$ vs time $t$, for a passive 3dKALJ model 
at temperature $T = 0.45$ for various system sizes in the range $N \in [400 - 100000]$.  
One can clearly see the huge increase in peak height with increasing system size. Note
that the time at which $\chi_4(t)$ peaks also increases with the system size. (b). 
Similar plot for active system with $f_0 = 2.5$ at temperature where $\tau_\alpha \sim 2200$
similar to the passive case. Notice the dramatic enhancement (nearly 10 fold) of the peak with 
increasing activity. (c,d) Variation of $\chi_4(t)$ peak with changing activity for 
$N = 10000$ (left) and $N  =  100000$ (right) particle system.}
\label{chi4p}
\end{center}
\end{figure*}
   
As reported in Ref.\cite{Mandal2016}, the degree of activity in the system can be 
quantified using a unique parameter $\Omega = c f_0^2 \tau_p$ and as long as $\Omega$
is same with various possible combinations of $c$, $f_0$ and $\tau_p$, the dynamical 
behaviour should be same. This is also known to be true over a small window of parameter
values.  In this study, we found that over the studied range of parameter values, this unique
activity parameter, $\Omega$ faithfully captures the effective degree of activity in the 
system. The system size dependence of $\chi_4^{P1}$ for canonical ensemble is very 
different from that of grand canonical ensemble as discussed before. In the main article
we have presented the data pertaining to the subsystems where all possible fluctuations 
can be included while measuring $\chi_4(t)$. The system size dependence of $\chi_4^{P1}$
when calculated for the full system is presented in Fig.\ref{FSS}(a) for
reference.
Note that dependence is very similar to that of the subsystems (blocks) and overall conclusions
do not change qualitatively even if one works with $\chi_4^{P1}$ for the full systems but there
are some issues that we observed while working with full system size data. Some of these
are elaborated in the subsequent sections. 

\section{Scaling Analysis of $\chi_4^{P1}$ using full system}
\begin{figure*}[!htpb]
\includegraphics[width=0.92\textwidth]{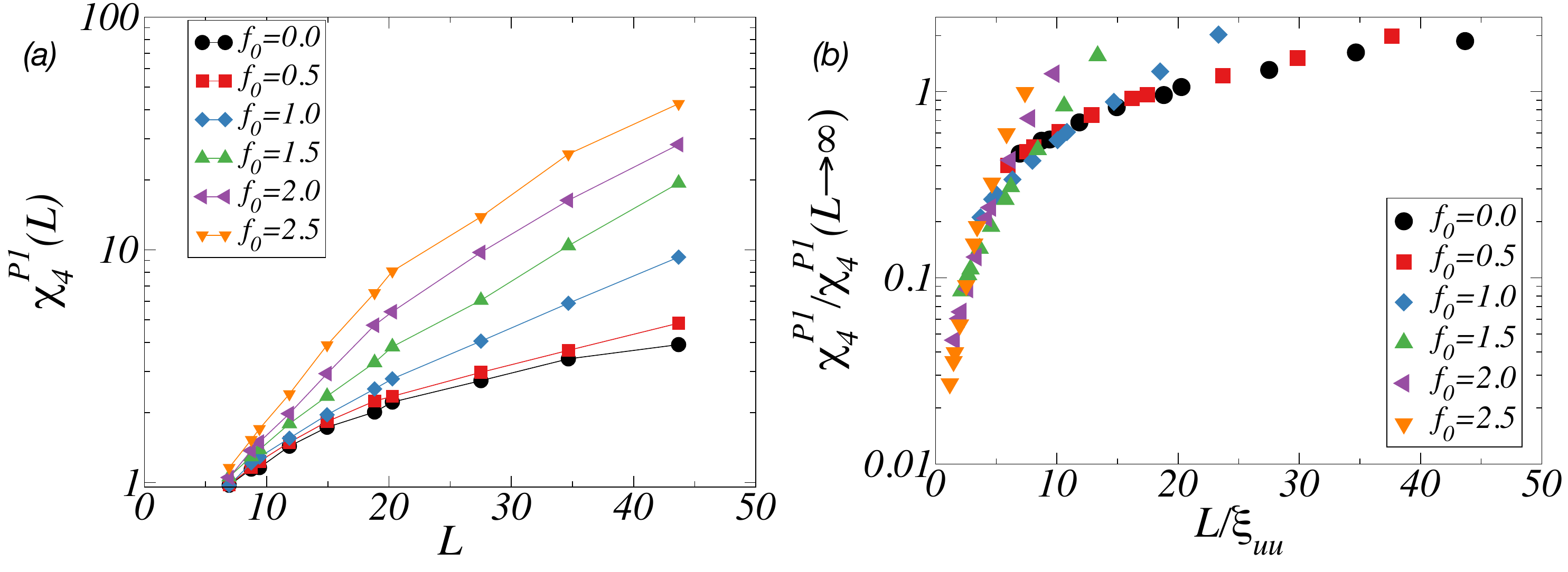}
\caption{(a) The system size dependence of $\chi_4^{P1}$ is plotted for systems with different active forces $f_0$ and the same concentration of active particles, $c=0.1$. (b) The finite system size data is tried to collapse using the length scale obtained from the displacement-displacement correlation function ($\xi_{uu}$). The data collapse obtained is not that great as presented for $L/4$ system size but does not dismiss the presence of the scaling ansatz. }
\label{FSS}
\end{figure*}
In the main text, we presented the finite-size data of $\chi_4^{P1}$, calculated for the quarter of the system length, i.e., $L/4$. Here in Fig.\ref{FSS}, we present the full system size data and the best scaling collapse possible by using the length scale obtained from $g^{uu}(r,t^*)$. One can see that the larger system size and larger activity data points are coming out of the collapse, which we infer are coming because of the missing fluctuations in the system as well as averaging issues, as mentioned in the main text. 
Once we take quarter of the systems for the doing the analysis (as 
shown in the main article) the issues disappear which can be due to two reasons. First being
that with quarter system size, one will have much better averaging as well as it is going to include
all possible missing fluctuations that are important in $\chi_4(t)$. The second reasons can be 
that the system size itself became smaller and if one does simulations of much bigger system 
sizes then one will again see the same deviation. At this moment we are constrained by the
larger system size ($N = 100000$) in our hand due to computational expense, so ruling out
the second possibility is not possible at this moment. The important part of this analysis is 
that the scaling ansatz still seems to be quite good to describe most part of the data.

\begin{figure*}[!htpb]
\includegraphics[width=0.98\textwidth]{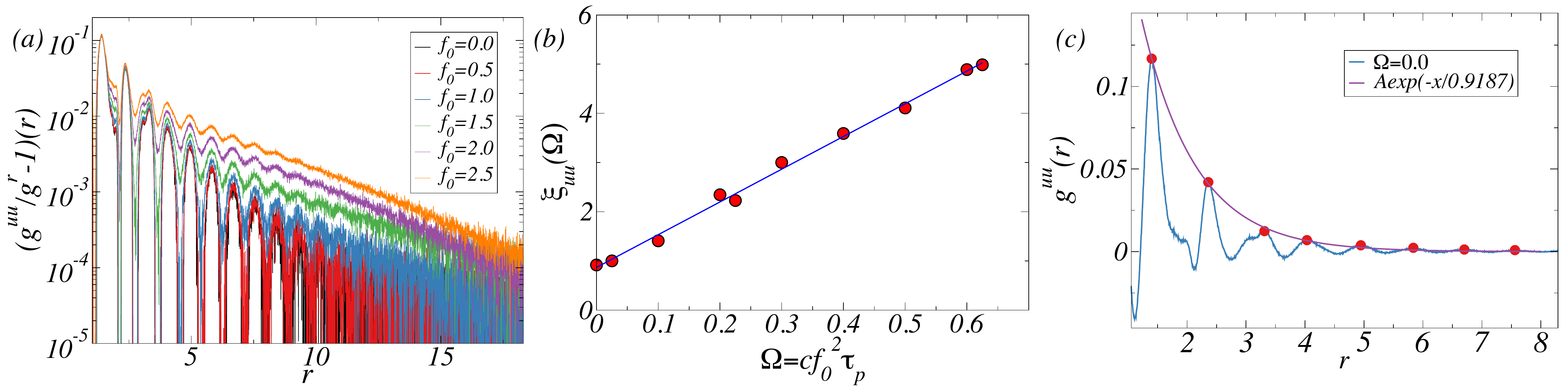}
\caption{(a)The quantity $(g^{uu}-g^r)/g^r(r,\Delta t)$ is plotted as a function of spatial distance for $\Delta t=t^*$, here $g^{uu}(r,\Delta t)$ is the displacement-displacement correlation function (Eq.\ref{guu}), and $g^r$ is the pair correlation function. The plotted quantity would exponentially decays to zero providing us the correlation length $\xi_{uu}$, which can be clearly seen increasing with increasing activity. The length scale variation with increasing activity can be obtained as a integrated area of the plots and is shown in panel (b). In panel (c) we fitted the peaks of the $g^{uu}(r)$ plot for zero activity ($\Omega=0$) with an exponential function to obtain $\xi_{uu}(0)$. This value is then used to scale the integrated areas for other activities. }
\label{guuFig}
\end{figure*}  
 \section{Displacement-displacement correlation function, $g^{uu}(r,t)$}
The dynamical length scale of the system $\xi_d$ can be computed independently 
by computing the displacement-displacement correlation function $g^{uu}(r,t^*)$ at 
$t^*$ \cite{TahPRR,Poole1998}. It is defined as,
\begin{equation}
g^{uu}(r,\Delta t)=\frac{\left\langle  \sum\limits_{i,j=1,j\neq i}^Nu_i(0,\Delta t)u_j(0,\Delta t)\delta(r-|\textbf{r}_{ij}(0)|)\right\rangle}{4\pi r^2\Delta rN\rho \langle u(\Delta t)\rangle^2}
\label{guu}
\end{equation}  
 where, $u_i(t,\Delta t)=|\textbf{r}_i(t+\Delta t)-\textbf{r}_i(t)|$, and $\langle u^2(\Delta t)\rangle=\langle\frac{1}{N}\sum_{i=1}^N
 \textbf{u}_i(t,\Delta t).\textbf{u}_i(t,\Delta t)\rangle$. $g^{uu}(r,\Delta t)$ is calculated at time $\Delta t=t^*$, along with the usual pair 
 correlation function $g^r(r)$ defined as,
 \begin{equation}
 g^r(r)=\frac{\left\langle\sum\limits_{i,j=1,j\neq i}^N\delta(r-|\textbf{r}_{ij}(0)|) \right \rangle} {4\pi r^2\Delta r N\rho}
\end{equation}  
 
 For far enough particles the displacement over a large enough time duration would be decorrelated and 
 $g^{uu}$ would be equal to $g^r$. So the quantity $g^{uu}(r,\Delta t)/g^r-1.0$ would decay to zero as a function of $r$.  If one 
 assumes the decay to be exponential, then the area under the curve would provide us with the correlation 
length.  Fig.\ref{guuFig}(a) contains the semi-log plot of $g^{uu}(r,\Delta t)/g^r-1.0$ calculated 
for system size of $N=10^5$. The obtained length scale ($\xi_{uu}$) with increasing activity in the system is plotted in 
Fig.\ref{guuFig}(b). The values in the plot are the integrated areas scaled by the value of the 
length scale obtained by fitting the data to an exponential function for the passive case, 
$f_0=0.0$ (Fig. \ref{guuFig}(c)).

\bibliographystyle{apsrev4-2}
\bibliography{Actphonon_SI}